# Human perception of audio deepfakes: the role of language and speaking style

Eugenia San Segundo[1]*, Aurora López-Jareño[1], Xin Wang[2], Junichi Yamagishi[2]

[1] Phonetics Laboratory, Spanish National Research Council, Madrid, Spain

[2] National Institute of Informatics, Tokyo, Japan

*Corresponding author

E-mail: eugenia.sansegundo@csic.es

# Abstract

Audio deepfakes have reached a level of realism that makes it increasingly difficult to distinguish between human and artificial voices, which poses risks such as identity theft or spread of disinformation. Despite these concerns, research on humans' ability to identify deepfakes is limited, with most studies focusing on English and very few exploring the reasons behind listeners' perceptual decisions. This study addresses this gap through a perceptual experiment in which 54 listeners (28 native Spanish speakers and 26 native Japanese speakers) classified voices as natural or synthetic, and justified their choices. The experiment included 80 stimuli (50% artificial), organized according to three variables: language (Spanish/Japanese), speech style (audiobooks/interviews), and familiarity with the voice (familiar/unfamiliar). The goal was to examine how these variables influence detection and to analyze qualitatively the reasoning behind listeners' perceptual decisions. Results indicate an average accuracy of 59.11%, with higher performance on authentic samples. Judgments of vocal naturalness rely on a combination of linguistic and non-linguistic cues. Comparing Japanese and Spanish listeners, our qualitative analysis further reveals both shared cues and notable cross-linguistic differences in how listeners conceptualize the "humanness" of speech. Overall, participants relied primarily on suprasegmental and higher-level or extralinguistic characteristics—such as intonation, rhythm, fluency, pauses, speed, breathing, and laughter—over segmental features. These findings underscore the complexity of human perceptual strategies in distinguishing natural from artificial speech and align partly with prior research emphasizing the importance of prosody and phenomena typical of spontaneous speech, such as disfluencies.

# 1. Introduction

A recent European policy report (van Huijstee et al., 2021, page I) defines deepfakes as "manipulated or synthetic audio or visual media that seem authentic, and which feature people that appear to say or do something they have never said or done, produced using artificial intelligence techniques, including machine learning and deep learning." Deepfakes may be conceptualized as a subcategory within the broader domain of AI-generated synthetic media, a domain that encompasses not only video and audio, but also still images and written text. However, our investigation focuses on only-audio deepfakes, which are samples that have been generated or modified by artificial intelligence to closely imitate genuine human voices (Khanjani et al., 2023).

Deepfake technologies can be employed for a wide variety of purposes, yielding both beneficial and harmful outcomes. On the positive side, they offer potential applications in domains such as audio-graphic productions, human–machine interactions (e.g., enhancing digital experiences), video conferencing, satire, personal or artistic creative expression, and medical research—for instance, the creation of synthetic voices for patients with neurodegenerative disorders. Although these applications highlight the constructive potential of synthetic media, deepfake technology also presents considerable societal risks. Individuals are increasingly exposed to misinformation and manipulation, as synthetic audio and video can be difficult to distinguish from authentic material. Moreover, deepfakes can undermine trust in audio-visual evidence and facilitate malicious practices such as fraud, identity theft, or reputational harm. Notably, both individuals and voice-recognition security systems are vulnerable to deepfake-based spoofing attacks (San

Segundo, 2023). According to van Huijstee et al. (2021), the risks associated with deepfake technologies can broadly be categorized into three types of harm: psychological, financial, and societal. On the individual level, targeted people may experience direct psychological effects. In addition, deepfakes can be produced and disseminated with the intention of generating financial damage. Finally, serious concerns remain about the broader societal implications of this technology: news media manipulation, damage to the justice system, the scientific system, and democracy, to name but a few.

Against this backdrop, it becomes crucial to improve detection technologies and increase public awareness of deepfakes and their potential dangers. Studying how humans perceive synthetic voices is particularly important because it helps to determine the extent to which people can be deceived. Perceptual research can also shed light on the linguistic and phonetic cues that differentiate authentic speech from its artificial counterpart. The current investigation seeks to advance the understanding of how well people can distinguish genuine voices from deepfakes in Spanish and Japanese, considering multiple factors that might influence detection accuracy. Specifically, we investigate:

1. Whether familiarity with the language affects the ability to tell deepfake voices from real ones.
2. Whether speaking style impacts the precision of voice discrimination.
3. Whether prior familiarity with a speaker's voice improves deepfake detection.

The key variables examined are language proficiency, speaking style, and voice familiarity. Past studies have reached mixed conclusions regarding language proficiency. For instance, Müller et al. (2022) reported that native English speakers detected English deepfake audio slightly better than non-native speakers, a finding also noted by Cooke et al. (2025). In contrast, Mai et al. (2023) found no significant differences in detection accuracy between native speakers of English and Mandarin.

As for speaking style, earlier perceptual studies—such as those by Mai et al. (2023) and Watson et al. (2022)—tended to rely on recordings of text being read aloud, rather than spontaneous speech. In this study, we address that gap by comparing speaking styles with varying levels of spontaneity. Spontaneous speech often contains features that are harder for AI models to replicate, including hesitations, filled pauses, and truncated or incomplete words, called disfluencies (McDougall & Duckworth, 2017). Consequently, deepfake interviews are likely to sound less natural due to the difficulty of imitating these spontaneous traits (Székely et al., 2019, 2020).

Additionally, evidence from related research supports the importance of speaking style in perceptual evaluations. Cooper & Yamagishi (2021), in a large-scale Mean Opinion Score (MOS) listening test, compared synthetic voices generated from different speaking styles—news, book sentences, and dialogues with a virtual assistant. Their listeners consistently rated news-style samples as the most natural, followed by the virtual-assistant dialogues, while audiobooks were judged significantly less natural. The authors suggest that the highly expressive style of many of the book sentences may appear unnatural when presented out of context. Although their task (MOS test) differs from ours (binary classification of natural vs. artificial voices), their findings reinforce the idea that speaking style strongly influences perceptual judgments.

With respect to voice familiarity, neuroscientific evidence suggests that familiar and unfamiliar natural voices are processed differently in the brain. For example, Bethmann et al. (2012)

observed stronger blood oxygenation level dependent (BOLD) signal amplitudes in the temporal lobes for familiar voices. However, few perceptual deepfake detection studies have addressed this variable. Mukhopadhyay et al. (2015) tested three familiarity conditions: unfamiliar voices, briefly familiarized voices (participants first listened to a short sample), and famous voices. In their study, the synthetic voices were generated with a Gaussian Mixture Model (GMM). Results showed no meaningful differences across conditions, with accuracy rates remaining around 50%. Wenger et al. (2021) replicated this experiment using a Text-to-Speech (TTS) cloning system. They obtained different results: while no differences emerged between unfamiliar and briefly familiarized voices, accuracy rose to approximately 80% when the voices belonged to famous speakers. Building on these findings, the present study seeks to further examine the role of voice familiarity in deepfake detection, addressing a gap that remains largely unexplored in the literature.

In sum, the following hypotheses are proposed:

- Language familiarity improves deepfake detection accuracy.
- Deepfake interview samples are easier to identify than audiobook samples.
- Familiar voices are more readily identified as deepfakes than unfamiliar ones.

In a second part of the study, we conducted a qualitative thematic analysis aimed at understanding the reasoning behind listeners' decisions, following similar methodologies to those implemented by Mai et al. (2023) and Warren et al. (2024). Mai et al. (2023), analyzed participants' open-ended responses using tf-idf (term frequency – inverse document frequency) weighting and generating word clouds to highlight salient lexical patterns. Their analysis revealed that participants often referred to the same features—such as pauses, tone, and intonation—regardless of whether their judgments were correct, suggesting that integrating such human-reported cues into automated detection systems would yield limited improvements. Both English- and Mandarin-speaking participants frequently relied on notions of "naturalness" and "robotic" qualities, while also referencing pauses, intonation, pronunciation, and speech rate. Cross-linguistic differences were observed: English participants more often mentioned breathing, whereas Mandarin participants emphasized cadence, word pacing, and fluency.

Interestingly, Warren et al. (2024) found that listeners correctly classify human audio at significantly higher rates than machine learning models and rely on linguistic features (together with other factors) when performing classification. Their qualitative analysis of listeners' judgments highlights key decision factors—including prosody, accents, and background noise—and reveals that the influence of these features differs across real and synthetic stimuli. A comparison with machine learning models shows that performance is not uniformly superior on the part of models; rather, humans and algorithms rely on distinct classification strategies.

The remainder of this paper is organized as follows: Section 2 describes the methodology, Section 3 presents the results, Section 4 discusses the findings, and Section 5 outlines our conclusions.

# 2. Materials and methods

## 2.1. Survey design

Using the open-source software PsychoPy (Peirce et al., 2019), we designed a phonetic perceptual experiment lasting approximately 25 minutes, which was then uploaded to Pavlovia for online distribution. The experiment could be completed on various electronic devices (PC, tablets, smartphones) without any time limitation. It aimed to assess human ability to distinguish between real human voices and their deepfake counterparts. Thus, participants were exposed to 80 audio clip stimuli presented in a randomized order, with an equal distribution of natural and artificial clips (one synthetic stimulus per counterpart: 40 natural and 40 deepfake). After listening to each clip, participants were required to classify it as either “natural,” if they believed it was a real human voice, or “artificial,” if they thought it was AI-generated. Note that the term “deepfake” was avoided, as it has been associated (Satariano & Mozur, 2023, in Warren et al., 2024) with a degree of stigma that may influence how individuals perceive or engage with media.

Additionally, after each response, they rated their confidence level on a 5-point Likert scale, where 1 indicated ‘not at all confident’ and 5 indicated ‘completely sure.’ Participants did not receive any feedback on their performance during the experiment, nor were they informed about the proportion of artificial to natural stimuli.

The experiment was divided into two parts, with a break in between. In the first part, the clips were extracted from audiobooks read in Spanish and Japanese. In the second part, the clips featured Spanish and Japanese celebrities speaking in interviews. This second part included two additional tasks: a yes/no question about whether they knew the celebrity’s voice and an open-ended question asking them to explain the reasoning behind their ‘natural’/‘artificial’ classification.

Two test interfaces were designed: one in Spanish and one in Japanese. The instructions and questions shown on each survey were exactly the same, but designing the experiments in the native language of each listener cohort was deemed important to ensure that they understood the required task and that they could reply in their mother tongue.

## 2.2. Participants

The final dataset comprised 4,291 responses: 2,211 from 28 native Spanish listeners (50 % male and 50 % female), and 2,080 from 26 native Japanese speakers (73.08 % male and 26.92 % female). All of them were recruited via the Pavlovia server. The Spanish participants’ age ranged from 22 to 65 years (M = 30.9 years, SD = 10.34). The Japanese participants’ age ranged from 23 to 63 years (M = 44.31 years, SD = 9.01). Additional potentially relevant characteristics were collected, including Japanese language proficiency (rated on a Likert scale from ‘null’ to ‘high’) and advanced linguistic knowledge (self-reported academic background). The distribution of these characteristics was as follows:

- Spanish listeners: regarding linguistic expertise, 39.29% of Spanish participants reported having specific training; regarding Japanese proficiency, 24 participants reported no competence, 3 reported low competence, and 1 reported an intermediate level.
- Japanese listeners: none of the participants reported either linguistic expertise; regarding Spanish proficiency, all of the participants reported no competence, except for 1 who reported low competence.

Individuals with hearing impairments were excluded from the study. At the outset of the survey, participants were presented with an informed consent page outlining the nature of the task, the expected time commitment, data security measures, and their right to withdraw at any time. They were asked to complete the experiment in a quiet environment using headphones.

| | n | Mean age (sd) | Male (%) | Linguistic knowledge (%) |
|---|---|---|---|---|
| **Spanish listeners** | 28 | 30.89 (10.34) | 50 | 39.29 |
| **Japanese listeners** | 26 | 44.31 (9.01) | 73.08 | 0.00 |
| **All** | 54 | 37.35 (11.77) | 66.11 | 20.37 |

**Table 1.** Demographic characteristics and linguistic knowledge of: Spanish listeners, Japanese listeners, and overall sample.

## 2.3. Stimuli selection

### 2.3.1. Bonafide stimuli

To obtain 20 real text-reading samples, we sourced ten Spanish and ten Japanese audiobooks from LibriVox and YouTube. Each stimulus was extracted from a different audiobook read by a different speaker. Then, the software Praat (Boersma & Weenink, 2024) was used to trim a 10-second fragment from each full recording.

The 10 stimuli of Spanish celebrity interviews were obtained from the VoxCeleb-ESP corpus (Labrador et al., 2023). The selected Spanish celebrities represented a broad spectrum of public figures, including singers, journalists, television hosts, actors, athletes, and comedians. Additionally, we aimed to include celebrities from various Spanish regions to capture geographic accent diversity. On the other hand, the 10 stimuli from Japanese celebrity interviews were sourced from the EACELEB corpus (Caulley et al., 2022). In this case, all selected celebrities were either actors or singers, and regional accent diversity could not be ensured. For both corpora, we established the following exclusion criteria before selecting the stimuli:

(a) Presence of background noise or music
(b) Poor recording quality
(c) Interruptions by the interviewer or audience
(d) Insufficient material in the corpus to generate a cloned voice
(e) Presence of political and controversial content

The last criterion was included to minimize extraneous cues in the content, as the experiment aimed to evaluate phonetic characteristics that can be used to distinguish real voices from fake voices.

Gender balance was maintained, resulting in an equal distribution of male and female voices for both Spanish and Japanese audiobooks and interview stimuli.

### 2.3.2. Voice cloning

We ensured that each synthetic voice reproduced the exact same phrase as its natural counterpart. This allowed for a direct comparison between real and artificially generated voices, isolating perceptual differences to the phonetic characteristics themselves rather than linguistic content. Thus, transcriptions of the natural stimuli were necessary to generate deepfakes that precisely replicated the original audio clips. The automated transcription tool Whisper (Whisper Transcribe, n.d.) was utilized for this purpose. Subsequently, the transcriptions underwent a rigorous review and correction process by two qualified linguists.

To produce synthetic versions of the natural voices, we used ElevenLabs' Text-to-Speech (TTS) software (ElevenLabs, n.d.). We applied the "Eleven Multilingual v2" model with its default settings for Stability, Similarity, Style Exaggeration, and Speaker Boost, adhering to ElevenLabs' "Best Practices" guidelines (ElevenLabs, n.d.-b) and the ElevenLabs Prompt Guide for inputting both audio and text. For voice cloning, each target voice requires a training audio sample of 1 to 2 minutes in duration. These training audio samples were extracted from the same corpora detailed in Section 2.3.1, ensuring that the selected experimental stimuli were excluded.

After training the ElevenLabs software and obtaining accurate transcriptions for each natural audio clip, the generation of artificial voices was conducted through an iterative process. For each voice, a minimum of three cloned versions was produced. In certain instances, adjustments to the transcriptions were made during this process to enhance the naturalness of the generated voices, necessitating multiple iterations. Finally, a single deepfake was selected for each voice through a consensus-based approach involving at least three researchers or native-speaking collaborators.

## 2.4. Data analysis

All statistical analyses were conducted in RStudio (version 2025.5.1.513.3) (Posit team, 2025).

For the descriptive statistics, participants' overall mean accuracy rate (%) was calculated as well as participants' mean accuracy for each stimulus type.

For the inferential analysis, four separate generalized linear mixed models were fitted using the glmer function from the lme4 package (Bates et al., 2015), and the ggplot2 package was used for data visualization (Wickham, 2016).

In all models, the dependent variable was the probability of a correct answer (binary outcome: correct or incorrect). Responses from Spanish and Japanese participants were analyzed independently. For each group of participants, two separate regression models were designed:

- Model 1 included responses to all stimuli. It was modelled with fixed effects of AUTHENTICITY (natural vs. artificial), LANGUAGE (Spanish vs. Japanese), SPEAKING STYLE (audiobook vs. interview) and CONFIDENCE LEVEL (1-5).
- Model 2 focused on responses to interview stimuli, as familiarity with speakers' voice was only assessed for interviews. This model included the fixed effects of AUTHENTICITY, CONFIDENCE LEVEL, and FAMILIARITY.

Initially, speaker's sex, participants' gender and linguistics knowledge were added as additional fixed predictors. However, as they were not significant in any case, they were excluded from the final models reported in the Result section. We also did not include the effect of age due to the uneven distribution of participants across age groups, which likely limits the statistical power to detect age-related effects. Furthermore, while we aimed to avoid an overcomplicated and uninterpretable structure, we incorporated the authenticity x language x speaking style interaction, because the marginal $R^2$ improved when it was included.

In both models, listeners and audio clips were incorporated as random intercepts to account for individual variability in performance and stimulus-related variability in difficulty.

To assess the explanatory power of the generalized linear mixed-effects models, we computed the coefficient of determination ($R^2$), using the *r2_nakagawa* function from the *performance* R package. This provided the marginal $R^2$, which reflects the variance explained by the fixed effects alone, and the conditional $R^2$, which reflects the variance explained by both fixed and random effects. Additionally, visual inspection of residual plots using *DHARMa* package (Hartig, 2024) confirmed no obvious deviations from homoscedasticity.

As part of our study, participants were asked to provide justifications for their classification decisions. This makes up the qualitative part of our investigation. A total of 1,760 responses (free text) were collected and manually categorized into thematic groups. In contrast with previous studies such as Warren et al. (2024), which developed a codebook through a discussion of common ideas and generated a number of keywords grouped together by likeness, in this investigation we have considered a different approach, based on the idea of componentiality of speech, which is a highly relevant concept in a number of phonetic investigations. The componentiality of speech means that speech can be divided into distinct components or layers. As explained by Leemann et al. (2024), forensic speech experts rely on this concept and examine a broad sample of these features to build the most comprehensive profile of a speaker possible.

In order to classify the judgments provided by the listeners in their qualitative responses, in our investigation we have deemed useful to follow the speech classification proposed by Leemann et al. (2024), which distinguishes three main groups: (1) segments, (2) suprasegmentals and (3) higher-level linguistic features. Besides, for the coding of disfluencies highlighted by the listeners, we have followed the taxonomy of disfluency types (TOFFA) proposed by McDougall and Duckworth (2017).

The free-form text responses of the listeners were reduced into single nouns and adjectives corresponding as closely as possible to phonetic concepts. The 20 most common words were depicted in a word cloud following Mai et al. (2023).

# 3. Results

## 3.1. Participants' Performance

The overall mean accuracy rate across the 54 participants was 59.11% (SD = 7.37%). However, when the groups are considered separately, Spanish listeners performed somewhat better than Japanese listeners, with mean accuracy rates of 60.52% and 57.60%, respectively.

As shown in Table 2, both groups generally obtained higher scores when classifying natural than artificial audios, although this effect was more pronounced in Japanese participants. For instance, in Japanese listeners, the difference between natural and artificial stimuli exceeded 30% in Japanese interviews. In contrast, for Spanish listeners the gap was smaller, never exceeding 16%.

Regarding speaking style, Spanish participants tended to classify interviews more accurately than audiobooks. In contrast, Japanese participants did not exhibit a consistent pattern across speaking style.

In terms of variability, Spanish participants' mean accuracy varied between 48.93% and 77.50%, while Japanese participants exhibited more extreme values, ranging from 36.92% to 80.38%. Notably, Japanese participants' accuracy was remarkably low for Spanish artificial stimuli: 36.92% in audiobooks and 38.85% in interviews.

For both groups, the highest accuracy rates achieved were observed for natural interviews in their native language: 77.5 % for Spanish listeners and 80.38% for Japanese listeners.

It should be noted that the standard deviations were relatively large, in some cases exceeding 20, indicating substantial variability in participants' performance.

**Table 2.** Mean % correct answers (and standard deviation) per group of listeners. Responses classified by speaking style, language, and nature of the audio: natural (N) and artificial (A).

| | | | | | | | | |
|---|---|---|---|---|---|---|---|---|
| **Spanish listeners** | **Overall** | | | | | | | |
| | 60.52 (8.20) | | | | | | | |
| | **Audiobooks** | | | | **Interviews** | | | |
| | 55.09 (8.26) | | | | 66.02 (10.60) | | | |
| | **Japanese** | | **Spanish** | | **Japanese** | | **Spanish** | |
| | 52.68 (11.10) | | 57.50 (10.58) | | 57.68 (15.12) | | 74.46 (12.20) | |
| | **N** | **A** | **N** | **A** | **N** | **A** | **N** | **A** |
| | 51.79 (17.65) | 53.57 (18.29) | 62.57 (12.83) | 51.43 (21.55) | 65.54 (20.52) | 48.93 (22.33) | 77.5 (17.35) | 71.07 (15.71) |
| **Japanese listeners** | **Overall** | | | | | | | |
| | 57.60 (6.16) | | | | | | | |
| | **Audiobooks** | | | | **Interviews** | | | |
| | 58.27 (7.96) | | | | 56.92 (8.26) | | | |
| | **Japanese** | | **Spanish** | | **Japanese** | | **Spanish** | |
| | 65 (12) | | 51.54 (9.25) | | 64.62 (13.56) | | 49.23 (9.13) | |
| | **N** | **A** | **N** | **A** | **N** | **A** | **N** | **A** |
| | 64.62 (21.21) | 65.38 (25.49) | 66.15 (16.75) | 36.92 (18.28) | 80.38 (21.26) | 48.85 (27.03) | 59.62 (22.18) | 38.85 (18.18) |

## 3.2. Hypotheses testing

As described in section 2.4, four regression models were estimated to test the three hypotheses of the study. Table 3 reports the variance explained by each model, with "S" indicating Spanish listeners and "J" indicating Japanese listeners. The variance explained by the fixed effects ranges between models from 7.9% to 15.8%.

**Table 3**. Total dependent variable's variance accounted by each regression model.

| | **Model 1S** | **Model 1J** | **Model 2S** | **Model 2J** |
|---|---|---|---|---|
| **Conditional $R^2$** | 0.184 | 0.229 | 0.197 | 0.274 |
| **Marginal $R^2$** | 0.079 | 0.113 | 0.151 | 0.158 |

## 3.2.1. Japanese listeners

The **language** of the stimuli had a significant main effect for artificial audiobooks ($p< .001$; Table 4A) and natural interviews ($p < .01$; Table 4C). In both cases, performance was significantly better when the stimuli were in the listeners' native language, supporting Hypothesis 1. However, language did not significantly affect natural audiobooks or artificial interviews.

**Speaking style** significantly influenced the probability of correct classification for Japanese stimuli, both natural and artificial, but in opposite directions depending on authenticity. Natural Japanese stimuli were more likely to be correctly identified when presented as interviews rather than audiobooks ($p < .05$), whereas artificial Japanese stimuli were more likely to be correctly identified when presented as audiobooks than as interviews ($p < .05$). By contrast, speaking style did not significantly affect classification of Spanish stimuli, regardless of authenticity.

**Authenticity** significantly influenced correct classification in all cases except Japanese audiobooks. Natural stimuli were more likely to be correctly identified than artificial ones for Spanish audiobooks ($p < .001$; Table 4A), Spanish interviews ($p < .05$; Table 4C), and Japanese interviews ($p < .001$; Table 4B).

Furthermore, higher **confidence levels** were associated with greater accuracy, particularly at confidence level 4 ($\beta = 0.39$, $p < .05$) and level 5 ($\beta = 0.57$, $p < .05$).

Figure 1 and Figure 2 provide a visual representation of these results.

On the other hand, the generalized linear mixed model 2J revealed that **voice familiarity** did not significantly influence classification performance**.**

**Table 4.** Output of the regression model 1 based on Japanese listeners' responses (model 1J).

**(A) Reference: artificial Spanish audiobooks**

| | β | SE | z value | p |
|---|---|---|---|---|
| **Intercept** | -0.87 | 0.29 | -3.03 | |
| **Authenticity: Natural** | 1.38 | 0.37 | 3.77 | <.001 *** |
| **Language: Japanese** | 1.25 | 0.37 | 3.40 | <.001 *** |
| **Speaking Style: Interview** | 0.12 | 0.36 | 0.32 | 0.747 |

**(B) Reference: artificial Japanese interviews**

| | β | SE | z value | p |
|---|---|---|---|---|
| **Intercept** | -0.36 | 0.29 | -1.26 | |
| **Authenticity: Natural** | 1.59 | 0.37 | 4.27 | <.001 *** |
| **Language: Spanish** | -0.39 | 0.36 | -1.09 | .276 |
| **Speaking Style: Audiobook** | 0.74 | 0.36 | 2.04 | .041* |

**(C) Reference: natural Spanish interviews**

| | β | SE | z value | p |
|---|---|---|---|---|
| **Intercept** | 0.17 | 0.28 | 0.60 | |
| **Authenticity: Artificial** | -0.92 | 0.36 | -2.55 | .011 * |
| **Language: Japanese** | 1.06 | 0.38 | 2.83 | .005 ** |
| **Speaking Style: Audiobook** | 0.34 | 0.36 | 0.94 | 0.347 |

**(D) Reference: natural Japanese audiobooks**

| | β | SE | z value | p |
|---|---|---|---|---|
| **Intercept** | 0.37 | 0.29 | 1.26 | |
| **Authenticity: Artificial** | 0.01 | 0.37 | 0.02 | .984 |
| **Language: Spanish** | 0.14 | 0.37 | 0.38 | .701 |
| **Speaking Style: Interview** | 0.86 | 0.38 | 2.28 | .023 * |

Note: Only fixed effects of authenticity, language and speaking style are reported. In each table (4A-AD), the reference condition varies. This allows us to examine the main effects variables of interest in all their combinations.

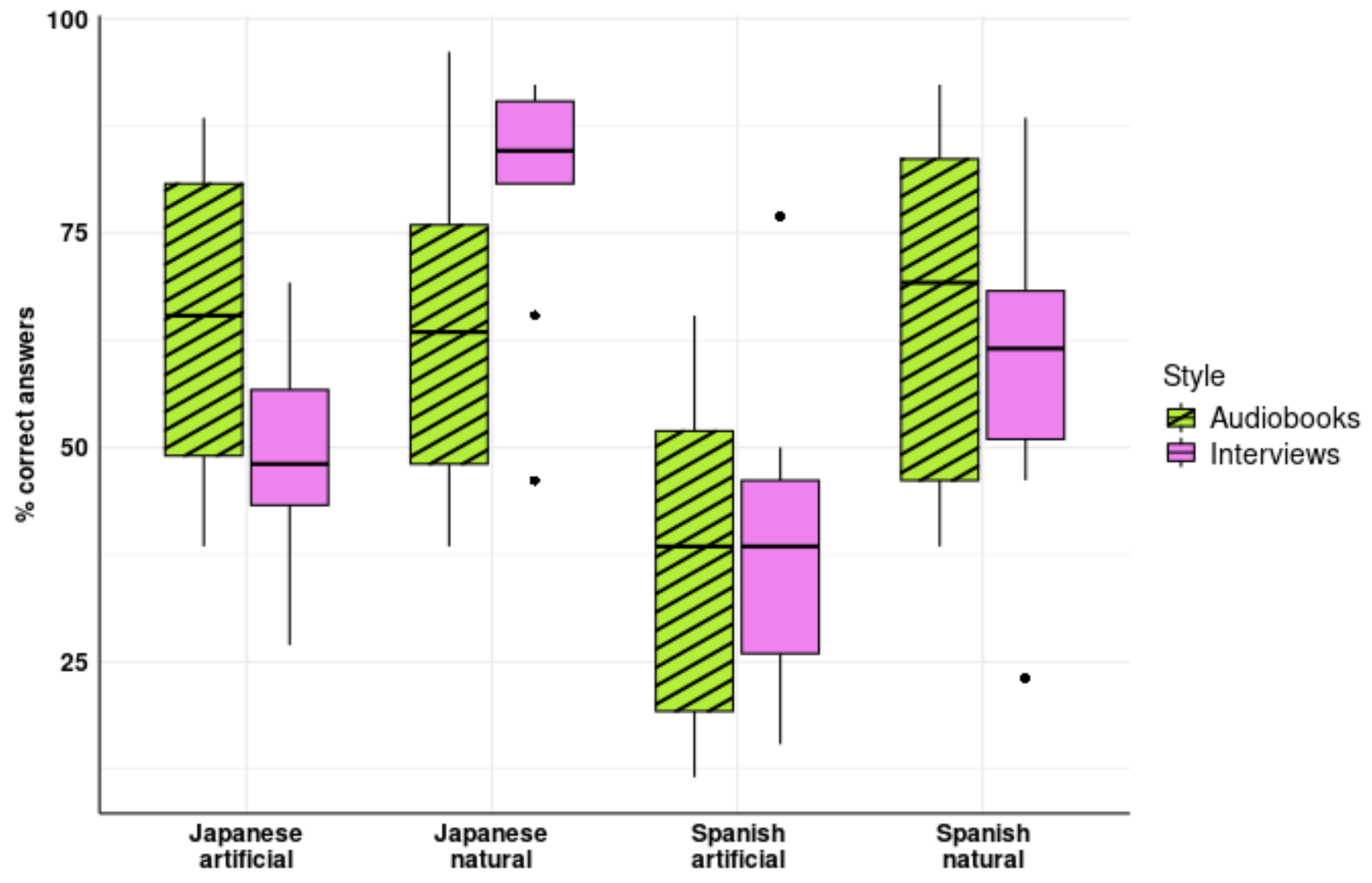

**Figure 1.** Japanese listeners' performance. Box plots of mean correct answer rate (%) per speaking style (audiobooks vs. interviews), audio language (Spanish vs. Japanese), and authenticity (natural vs. artificial).

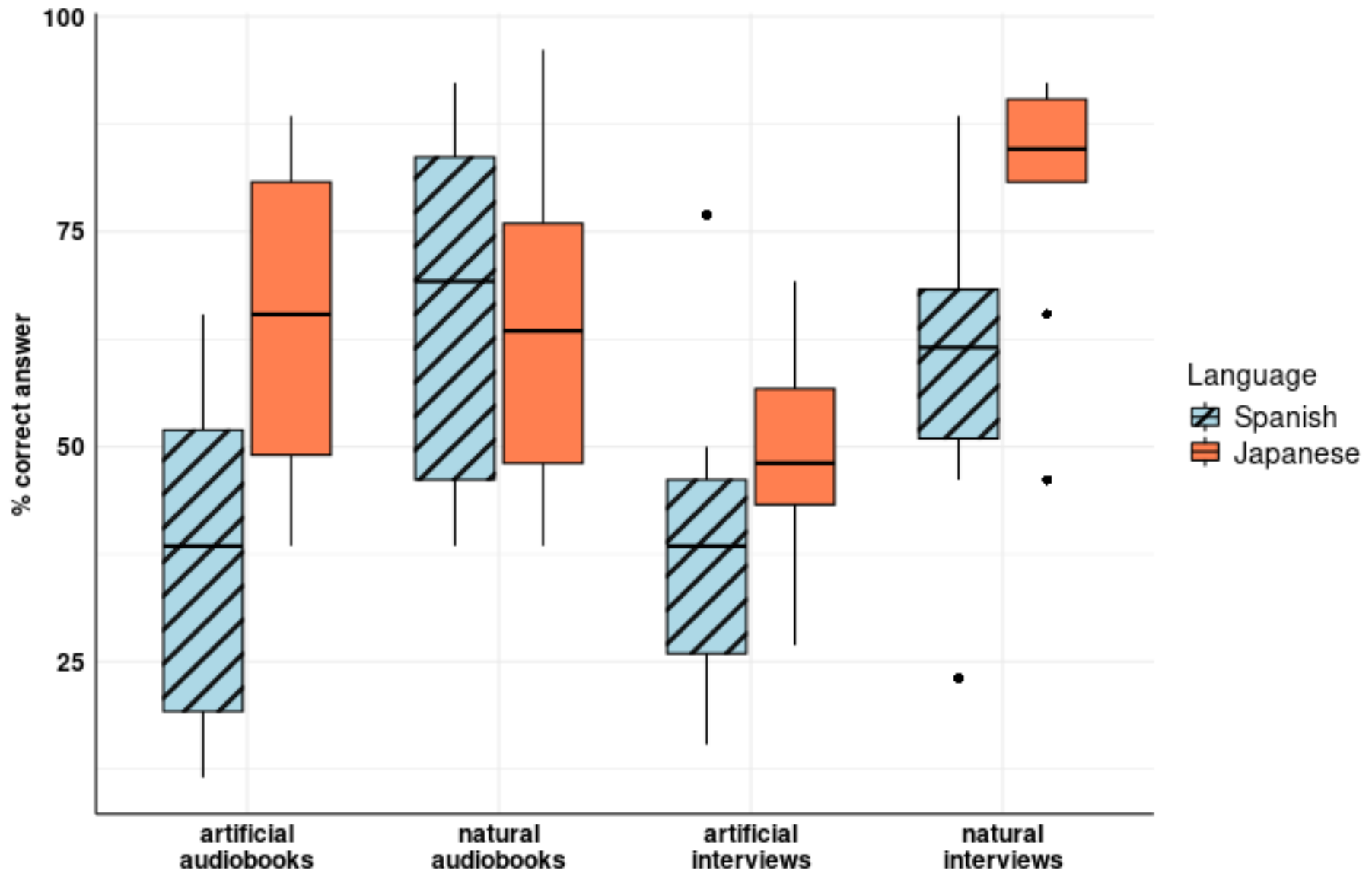

**Figure 2.** Japanese listeners' performance. Box plots of mean correct answer rate (%) per audio language (Spanish vs. Japanese), speaking style (audiobooks vs. interviews) and authenticity (natural vs. artificial).

### 3.2.2. Spanish listeners

The effect of **language** reached statistical significance for the artificial interviews ($p < .05$), with Spanish listeners performing more accurately on Spanish than on Japanese artificial interviews (Table 5B), whereas it did not have significant effect on the rest of conditions.

Regarding **speaking style**, it significantly affected the classification of Spanish stimuli. Spanish interviews were more likely to be correctly identified than Spanish audiobooks, both for artificial stimuli ($p < .01$; Table 5A) and for natural stimuli ($p < .05$; Table 5C). However, speaking style had no significant effect on Japanese stimuli.

**Authenticity** significantly influenced classification for Japanese interviews ($p < .05$; Table 5B), with natural interviews being more likely to be correctly identified than artificial ones. In contrast, authenticity did not significantly affect the classification of audiobooks (in either language) or Spanish interviews.

Additionally, as in Japanese participants, higher confidence levels were positively associated with accuracy, particularly at level 4 ($\beta = 0.96$, $p < .01$) and level 5 ($\beta = 1.49$, $p < .001$).

See Figure 3 and Figure 4 for a visual representation of these results.

**Table 5.** Output of the regression model 1 based on Spanish listeners' responses (model 1S). Only fixed effects of authenticity, language and speaking style are reported.

**(A) Intercept: artificial Spanish audiobooks**

| | β | SE | z value | p |
|---|---|---|---|---|
| **Intercept** | -0.75 | 0.41 | -1.84 | |
| **Authenticity: Natural** | 0.58 | 0.32 | 1.80 | .072 |
| **Language: Japanese** | 0.22 | 0.32 | 0.71 | .477 |
| **Speaking Style: interview** | 0.87 | 0.32 | 2.71 | .007 ** |

**(B) Intercept: artificial Japanese interviews**

| | β | SE | z value | p |
|---|---|---|---|---|
| **Intercept** | -0.67 | 0.40 | -1.66 | |
| **Authenticity: Natural** | 0.63 | 0.32 | 1.98 | .048 * |
| **Language: Spanish** | 0.77 | 0.32 | 2.39 | .017 * |
| **Speaking Style: Audiobook** | 0.12 | 0.31 | 0.40 | .692 |

**(C) Intercept: natural Spanish interviews**

| | β | SE | z value | p |
|---|---|---|---|---|
| **Intercept** | 0.49 | 0.42 | 1.17 | |
| **Authenticity: Artificial** | -0.39 | 0.33 | -1.16 | .245 |
| **Language: Japanese** | -0.53 | 0.33 | -1.60 | .110 |
| **Speaking Style: Audiobook** | -0.68 | 0.33 | -2.05 | .040 * |

**(D) Intercept: natural Japanese audiobooks**

| | β | SE | z value | p |
|---|---|---|---|---|
| **Intercept** | -0.63 | 0.41 | -1.54 | |
| **Authenticity: Artificial** | 0.09 | 0.31 | 0.27 | 0.786 |
| **Language: Spanish** | 0.44 | 0.32 | 1.36 | 0.173 |
| **Speaking Style: Interview** | 0.59 | 0.32 | 1.85 | 0.064 |

Note: Only fixed effects of authenticity, language and speaking style are reported. In each table (5A-AD), the reference condition varies. This allows us to examine the main effects variables of interest in all their combinations.

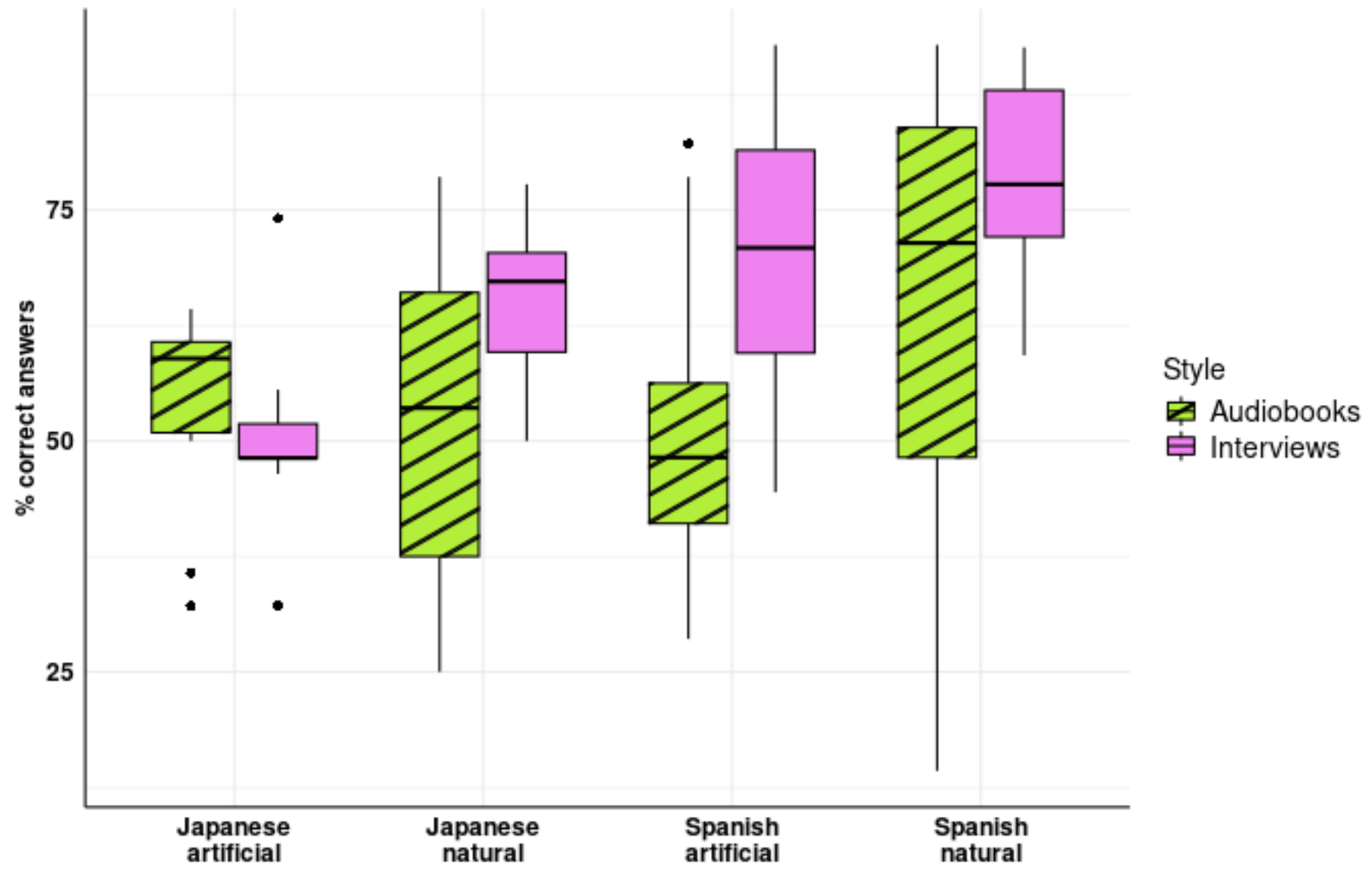

**Figure 3.** Spanish listeners' performance. Box plots of mean correct answer rate (%) per speaking style (audiobooks vs. interviews), audio language (Spanish vs. Japanese), and authenticity (natural vs. artificial).

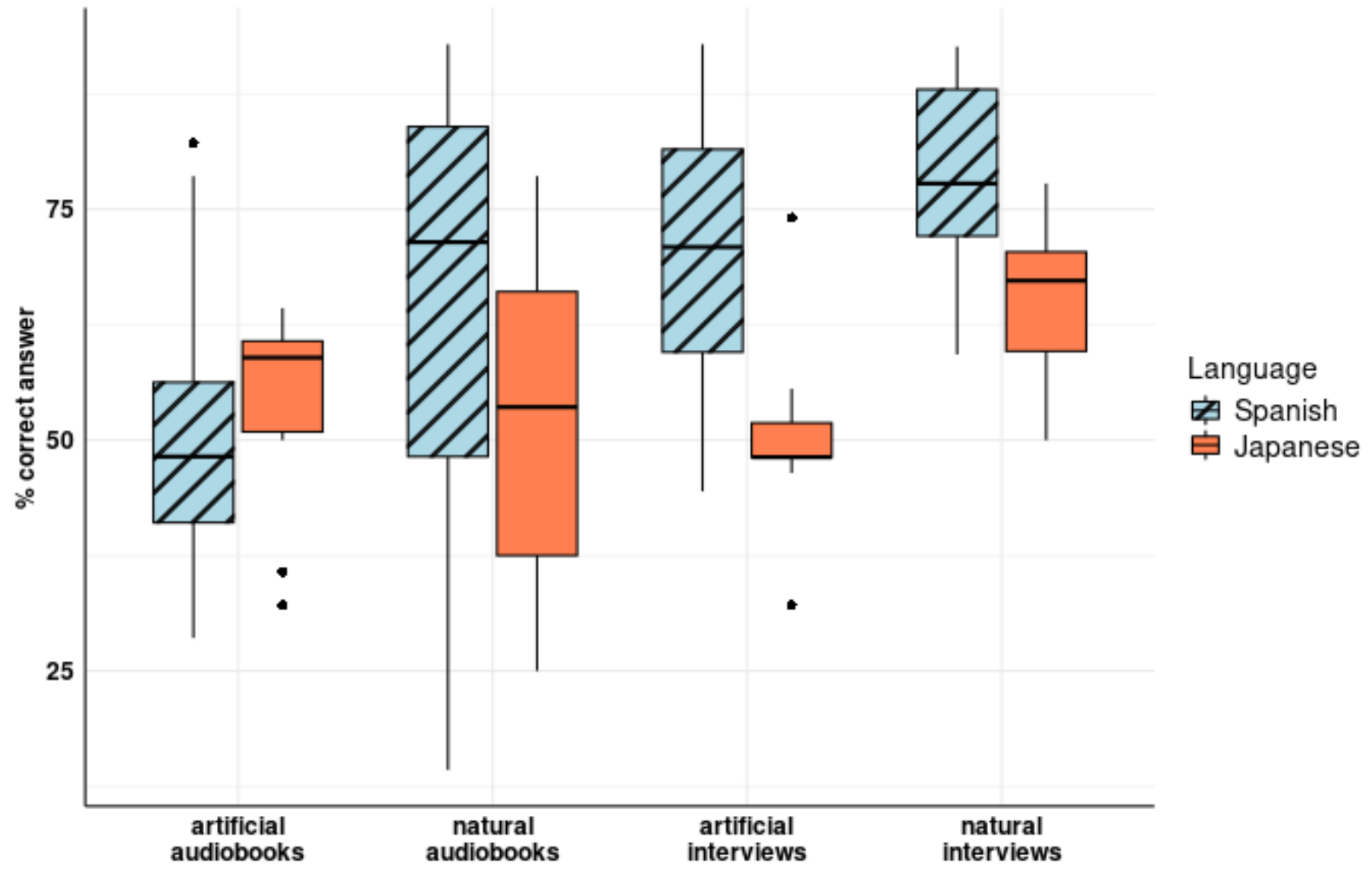

**Figure 4.** Spanish listeners' performance. Box plots of mean correct answer rate (%) per audio language (Spanish vs. Japanese), speaking style (audiobooks vs. interviews) and authenticity (natural vs. artificial).

## 3.3. Qualitative responses: What information do listeners attend to?

### 3.3.1. Japanese listeners

We input Voyant tools (Sinclair & Rockwell, 2016) with all the words derived from converting the qualitative responses of the listeners into single nouns and adjectives corresponding as closely as possible to phonetic concepts. The analysis revealed 678 total words and 59 unique word forms. Unique words (occurrence in brackets) were the following:

*inflection (61); intonation (44); fluency (44); way (41); tone (39); breathing (36); rhythm (30); pause (30); pitch (23); smooth (22); natural (21); mechanical (21); hesitation (21); muffled (18); emotion (17); laughter (15); speed (14); voice quality (12); discomfort (12); repetition (11); synthetic (10); sound quality (9); rephrasing (8); prolongation (7); filler (7); connected speech (7); electronic (6); accent (6); uncomfortable (5); machine-like (5); ending (5); unnatural (4); trembling (4); monotonous (4); intensity (4); emphasis (4); tongue click (4); voice (3); vocalization (3); unique (3); tension (3); stagnation (3); realistic (3); unrealistic (2); timbre (2); stuttering (2); stress (2); rough (2); reverberation (2); nonchalant (2); imperfection (2); gritty (2); formal (2); dynamics (2); clear (2); calm (2); volume (1); variation (1); unhuman (1)*

The 20 most common words are shown in a word cloud (Figure 5). Out of these 20 words, 16 are nouns while 4 are adjectives. The most common adjectives used by Japanese listeners are *smooth*, *natural* and *mechanical*, with 21 occurrences each. *Smooth* and *natural* are mostly used to describe natural stimuli while *mechanical* is used to refer to artificial stimuli, although we found also the utterance "lack of natural" followed by any of the phonetic aspects referred to in the nouns, mostly "lack of natural inflection" or "lack of natural intonation". The other adjective (*mechanical*) typically follows the same nouns, but a common collocate is also "rhythm". A fourth adjective is commonly used by Japanese listeners: *muffled.* They use it to refer to a certain voice quality, both in natural and artificial stimuli. Due to the subjective, synesthetic nature of this adjective, it is not clear what the listeners refer to. Commonly a muffled voice is a softened, muted, or little distinct voice, so it may refer to the phonetic concept of amplitude. Other adjectives tend to go together with 'voice quality' in the original responses, such as "smiling voice quality". For the sake of simplification, we reduced the full utterance to 'voice quality'.

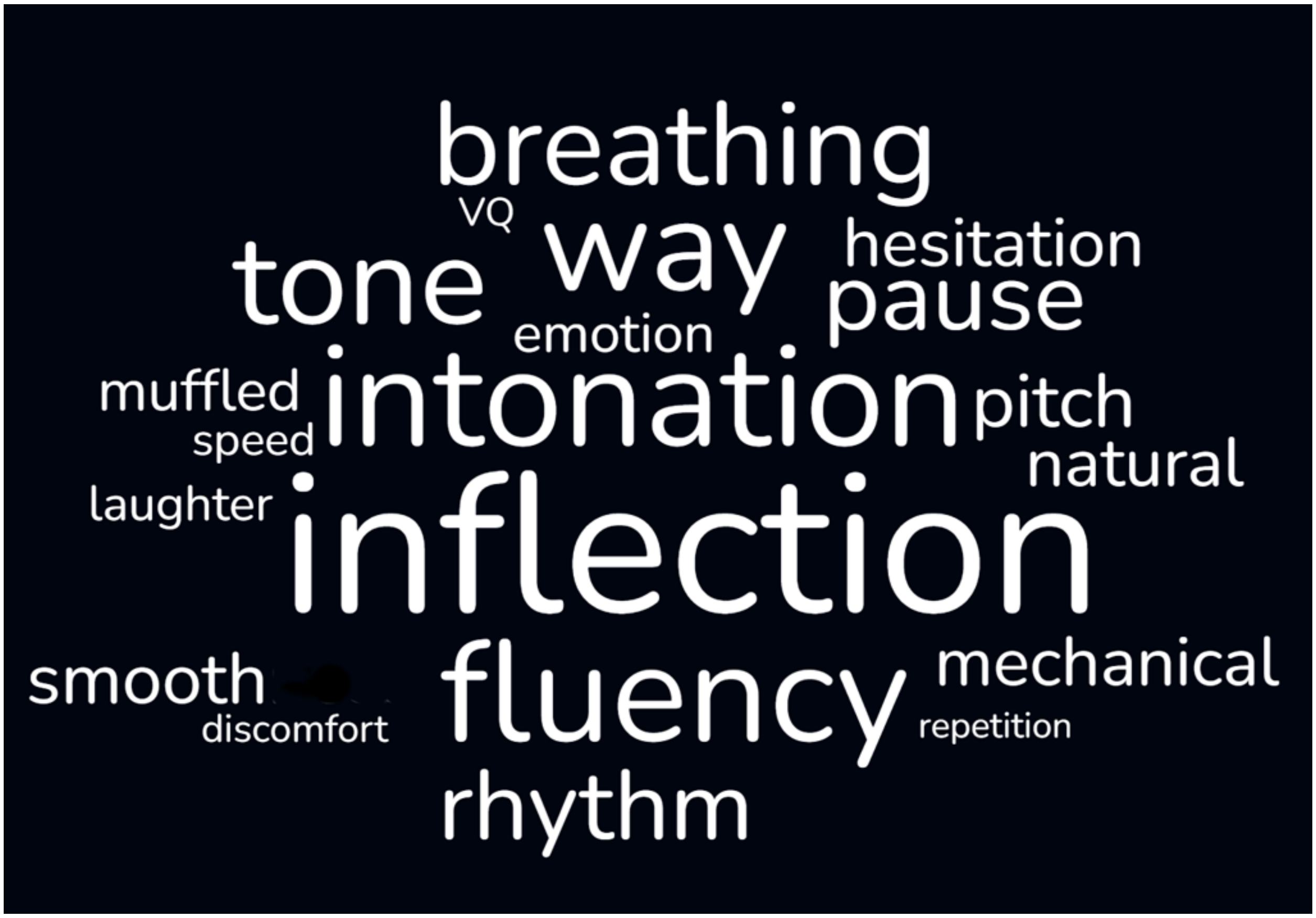

**Figure 5.** Word cloud with the 20 most frequent words found in the Japanese listeners' judgments (created with Simplewordcloud.com). VQ stands for 'voice quality'. Way stands for 'way of speaking'.

**Table 6**. Most frequent nouns (adjectives excluded) extracted from the 20 top frequent words found in the qualitative responses of Japanese listeners. Nouns are classified in segments, suprasegmentals and higher level with non-linguistic features.

| Segments | Suprasegmentals | Higher-level and non-linguistic features |
|---|---|---|
| N/A | Inflection | Fluency |
| | Intonation | Speaking way |
| | Tone | Breathing |
| | Rhythm | Pause behavior |
| | Pitch | Hesitation |
| | Speed | Emotion |
| | Voice quality | Laughter |
| | | Discomfort |
| | | Repetition |

The 16 most frequent nouns used by Japanese listeners are described in Table 1. Half of them can be classified within suprasegmentals; half of them within the higher-level and non-linguistic features. No nouns relate to segments.

In the group of **suprasegmentals**, we find references to an array of features that are not found at the local segmental level (i.e. segmental features that involve units of sound) but exist 'above' them or affect a series of segments. These are also called prosodic features:

- *Rhythm*, *pitch*, and *voice quality* are prosodic aspects that are highlighted by Japanese listeners as key aspects to decide when a stimulus was natural or artificial.
- *Speed* is the umbrella term that we used to cover any aspect in which listeners mention fast or slow articulation/speech rate.
- Regarding *tone*, the most probable meaning of this noun, often repeated by listeners, is, in general terms, the feeling conveyed by the way someone speaks. Most of the time, the expression used by the listeners is "tone of voice".
- In the case of *inflection*, there is reason to think that *inflection* and *intonation* are used by listeners to refer to the same idea (i.e. pitch fluctuation), which would considerably increase the occurrence of this phonetic concept in the listeners' explanations. However, we decided to leave both concepts for two reasons. On the one hand, most listeners used both words systematically and not interchangeably. On the other hand, inflection could also refer to amplitude fluctuation, and not only pitch fluctuation.

In terms of **higher level features and non-linguistic aspects**, we highlight the following aspects:

- *Fluency* appears with the highest frequency. Interestingly, fluency is associated with natural stimuli, but also disfluencies are associated with human voices. As explained by McDougall & Duckworth (2017), speakers interrupt the flow of their speech in different ways (pauses, sound prolongations, repetitions, self-interruptions, etc.). These phenomena are associated with speech planning and, although speakers exhibit large variation in their use and frequency of occurrence, participants of this experiment strongly associated with a human characteristic and thus with natural stimuli.
- *Speaking way* appears with the second highest frequency. This term was used to cover at least these alternatives: "way of speaking" or "manner of speaking". Actually, manner of speaking has been considered a category of its own by some forensic phoneticians (Künzel 1987, Jessen 2008; in Braun 2020) who distinguish three main feature categories in FSC: voice, speech and manner of speaking. For the above-mentioned authors, manner of speaking includes nonverbal features such as hesitations, question tags, breathing patterns, and clicking sounds. Since listeners tend to clearly identify breathing, laughter, etc. as isolated aspects that made them decide whether a stimulus was artificial or natural, we have let them as separated nouns, apart from the umbrella term "speaking way". We consider that "speaking way" is used by listeners to refer to what Leemann et al. (2024) describe as "other idiosyncratic higher- level features", including how speakers make use of their language variety's morphology, syntax, and lexicon. This idiosyncratic use could range from question tag particular choice, use of non-standard grammatical features or specific discourse patterns in conversation (e.g., the discourse marker *like*).

- *Breathing* and *laughter* are classical non-linguistic features analyzed by experts in FVC. They seem to have been also key for listeners to inform their decision about the nature of the stimuli in the perceptual experiment. *Laughter* appears less frequent than *breathing* since only a few natural audios contained laughter. Interestingly, breathing is the top non-linguistic feature mentioned by listeners, with 36 occurrences, more times than rhythm, pitch or voice quality, to mention a few.
- *Pause* is another key noun that is repeated frequently by listeners. Pauses, hesitations and repetitions belong to the group of disfluencies (McDougall and Duckworth, 2019). Disfluencies are high level features typical of spontaneous speech. These concepts appear associated with natural stimuli in Japanese listeners, but above all, *pauses* have the higher occurrence. If fillers, hesitation and prolongations had been considered together, as they typically refer to the same cognitive phenomenon of thinking while talking, their occurrence would be even higher.
- The word *emotion* typically appears in this type of perceptual studies, with emotions clearly associated with natural stimuli or human speech. "Lack of emotion" or "emotionless" is used to refer to artificial stimuli. Upon checking the translation from the original Japanese word (違和感), the noun *discomfort* seems to be used by many listeners as a synonym of unnatural, so it should probably be analyzed together with the three other frequent adjectives (*natural*, *mechanical* and *smooth*) used to describe stimuli.

Since our reduction of the listeners' judgements to a single noun or adjective could be sometimes too reductionist, we now include some full sentences that we found particularly insightful into what makes a voice human for naïve listeners. They all correspond to correct answers. For instance, the voice was artificial and the listener correctly selected the option 'artificial'. We deem that this type of qualitative responses will be useful to improve deepfake detection algorithms if we particularly consider the feedback associated with correct answers.

#1 *I thought it would be difficult to create the feeling of breathing when talking with synthesized sounds*. (Spanish, natural stimulus)
#2 *The connection between sentences was not human-like* (Spanish, artificial stimulus)
#3 *The way he spoke with an emphasis on onomatopoeic words such as "burn" and "garn" was very human.* (Japanese, natural stimulus)
#4 *It felt like the sound was choppy with each sentence, and it was very machine-sounding.* (Japanese, artificial stimulus)

## 3.3.2. Spanish listeners

We input Voyant tools with all the words derived from converting the qualitative responses given by the listeners into single nouns and adjectives corresponding as closely as possible to phonetic concepts. The analysis revealed 711 total words and 58 unique word forms. All the unique words (occurrence in brackets) were the following:

*intonation (130); pause (66); speed (53); emotion (46); hesitation (34); breathing (33); tone (29); timbre (25); flat (25); rhythm (22); prosody (16); laughter (15); intensity (14); inflection (14); consonant (13); fluency (11); emphasis (11); repetition (10); pitch (10); voice quality (9);*

*prolongation (9); variation (8); segments (8); metallic (8); stress (7); echo (7); geolect (6); connected speech (6); speaking way (5); trembling (5); robotic (5); monotonous (5); filler (5); background (5); accent (4); vivacious (3); theatrical (2); resonant (2); rephrasing (2); phoneme (2); nuances (2); ending (2); clipping (2); volume (1); speech quality (1); spontaneous (1); sigh (1); saliva (1); perfection (1); natural (1); informal (1); improvisation (1); hoarse (1); expressive (1); depth (1); click (1); clear (1); cadence (1).*

The 20 most common words are shown in a word cloud (Figure 2). Out of these 20 words, 19 are nouns while one is an adjective: *flat*. This adjective is used 25 times, typically to refer to intonation, which is the most common noun. Comparatively, Spanish listeners use less adjectives and more nouns. Many of the adjectives that they use are similar to the ones used by Japanese listeners: *monotonous* is used by both groups to refer to artificial stimuli. While Japanese prefer *synthetic*, *electronic* and *machine-like* to further characterize artificial voices, Spanish speakers often use *metallic* and *robotic*. In contrast, to refer to human voices, Japanese listeners almost exclusively use the adjective *natural.* For this purpose, Spanish listeners, while preferring to use nouns —and sometimes quite technical phonetic terms — rather than adjectives, when they use them, they resort to an array of adjectives, such as *vivacious*, *expressive*, or *resonant* voices.

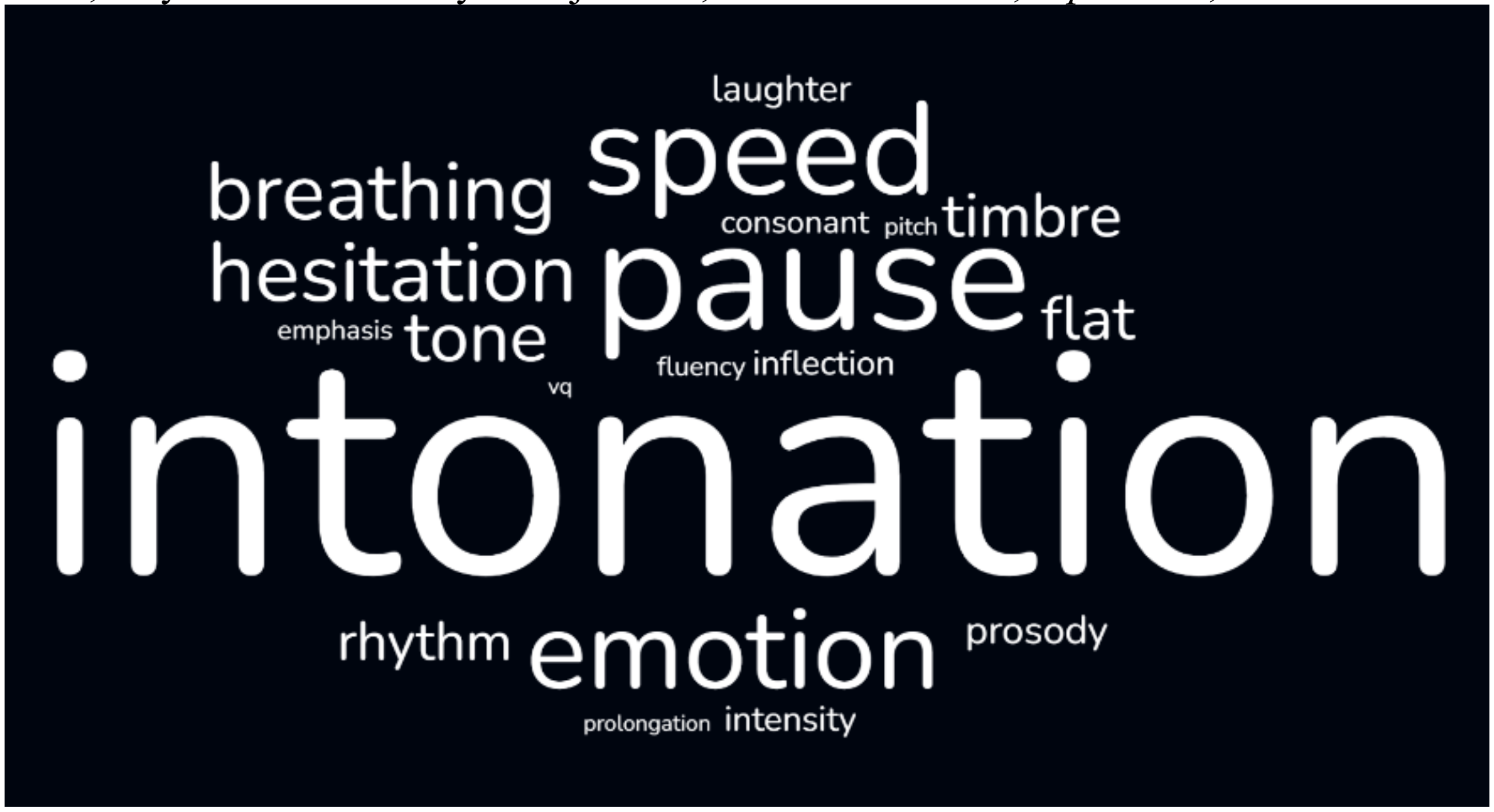

**Figure 6.** Word cloud with the 20 most frequent words found in the Spanish listeners' judgments (created with Simplewordcloud.com). VQ stands for 'voice quality'.

**Table 7.** Most frequent nouns (adjectives excluded) extracted from the 20 top frequent words found in the qualitative responses of Spanish listeners. Nouns are classified in segments, suprasegmentals and higher level with non-linguistic features.

| Segments | Suprasegmentals | Higher-level and non-linguistic features |
|---|---|---|
| Particular consonant | Intonation | Pause behavior |
| | Speed | Emotion |
| | Tone | Hesitation |
| | Timbre | Breathing |
| | Rhythm | Laughter |
| | Prosody | Fluency |
| | Intensity | Emphasis |
| | Inflection | Repetition |
| | Pitch | |
| | Voice quality | |

The top 20 words used by Spanish listeners include 19 nouns which were classified in three groups: segments, suprasegmentals and higher level with non-linguistic features (Table 2). In a similar way to what we found for Japanese listeners, most concepts fall within either the suprasegmentals group or the higher-lever and non-linguistic features. However, we also find 13 occasions in which a particular consonant is described by listeners as being the cue to make their choice on whether the stimulus was natural or artificial. This contrasts with the Japanese participants, who did not make any reference to segments. Following Leemann et al. (2024), 'segments' include consonants, vowels and connected speech processes. A possible explanation for this profusion of phonetic fine-grained knowledge found in the responses of Spanish listeners could be due to the fact that 39 % of the participants (11 subjects) reported having some kind of advanced linguistic knowledge. None of the Japanese participants, in contrast, reported having advanced linguistic knowledge, understood as being at least enrolled in an undergraduate degree in Linguistics at the moment of taking the perceptual test.

In terms of **segments**, it seems worth describing the 13 instances where a consonant was mentioned as the key for a participant to decide whether the audio was natural or artificial (Table 3).

**Table 8**. Segments (consonants) mentioned by listeners as cues to decide whether the voice was natural or artificial. Information about the specific audio (LANGUAGE-nature) and specific listener is provided, together with the full comment and information about whether the guess was correct or not.

| Consonant | Audio | Listener | Full comment | Correct response |
|---|---|---|---|---|
| /s/ | 47 (SPA-a) | 10 | Rosario doesn't pronounce her /s/ that much when she speaks and here they are overemphasized, so I think it's generated with AI. | YES |
| /s/ | 69 (SPA-a) | 10 | The pronunciation of the /s/ does not seem natural, but forced. | YES |
| (general) | 70 (JAP-a) | 10 | The pronunciation of the letters seems rushed, as if glued together. | YES |
| /r/ | 73 (SPA-n) | 10 | Unlike the previous audio, a much slower pace, hesitations and greater pronunciation of the /r/ | YES |
| (general) | 80 (JAP-n) | 10 | I appreciate a characteristic way of pronouncing some letters, there are pauses and different tones | YES |
| /s/ | 55 | 11 | The articulation of the /s/ | YES |
| | (SPA-n) | | | |
| /t/ | 67 (SPA-n) | 14 | She does not pronounce the voiceless dental occlusive at the end of 'comfort' | YES |
| /s/ | 41 (SPA-n) | 19 | Sound of /s/, fillers, word chaining | YES |
| /s/ | 65 (JAP-a) | 19 | Variation in the length of pauses. Sound of /s/ | NO |
| Sibilants | 46 (JAP-n) | 21 | Sibilants | YES |
| /f/ | 67 (ESP-n) | 21 | The /f/ in "fanática" | YES |
| /s/ | 71 (SPA-a) | 21 | Final /s/ | NO |
| /d/ | 76 (SPA-n) | 21 | /d/ sounds | YES |

In 84.62 % of the occasions, listeners guess correctly whether the voice is natural or artificial when they use segmental cues. Only 5 listeners use this type of information and sometimes their response is due to a combination of segmental and suprasegmental or high level characteristics. However, it seems worth resorting to this type of phonetic knowledge for making these perceptual decisions, provided that listeners have the auditory skills to detect the segments. It is remarkable that listeners also give correct answers even if the language is not their mother tongue. The two cases when they give incorrect answers, stimuli are artificial. This can be understood as a sign of an accurate cloning.

In the group of **suprasegmentals**, the most remarkable aspects are:

- Spanish listeners refer to almost the same concepts as those highlighted by Japanese. Yet, the frequency of occurrence of these nouns differs slightly. For instance, in the answers given by Spanish participants there are more references to *speed* than in the responses provided by the Japanese group.
- The term *prosody* was not used at all by the Japanese group but appears 16 times in the responses of the Spanish listeners. *Prosody* and *suprasegmentals* are typically used as synonyms. On the one hand, we do not know what the participants refer to when they use it, as it is rather an umbrella term: it could be intonation, rhythm, articulation rate. On the other hand, it is quite a technical name which a lay listener, with no linguistic knowledge, would rarely use.
- It is also remarkable that *timbre* is used more often by Spanish listeners. *Timbre* and *voice quality* are used indistinctly by phoneticians, the fact that timbre is the top four suprasegmental concept used by Spanish listeners is indicative of the high percentage of participants with a linguistic academic background among Spanish participants. Interestingly, when listeners use timbre, it is not followed by an adjective. It is just the general *timbre* (the distinctive quality of a voice) what makes listeners decide if an audio is natural or artificial. However, when they use *voice quality*, it is followed by common words used to describe an attribute of voice quality, or a type of voice quality (or setting, if we use Abercrombie and Laver's terminology). As an example, we find whisper or raspy voice quality, besides smiling VQ.

As far as the third group is concerned (**higher-level and non-linguistic features)**:

- We find almost exactly the same concepts used by Japanese, with different order of frequency. The only exceptions are: Japanese use more *speaking way* and Spanish use more *fluency*.
- *Discomfort* never appears in Spanish' responses, but the noun *emphasis* does appear instead.
- All in all, we say that both listener groups use the same disfluencies (hesitations, repetitions, pauses, etc.) to discriminate natural from artificial stimuli.
- Both groups pay attention to the presence of non-linguistic features such as *breathing* and *laughter* to decide that a voice is human.

Since our reduction of the listeners' judgements to a single noun or adjective could be sometimes too reductionist, in the same way as we did with the Japanese listeners, we have selected a few full sentences that we found particularly insightful into what makes a voice human for this group of listeners. All these comments are linked to correct answers, so they constitute particularly useful feedback for detecting artificial voices.

#1 *In the tone, timbre, and variety of Spanish he speaks. He uses Andalusian Spanish, and in this recording, Castilian Spanish is used.* (Spanish, artificial stimulus)
#2 *For the way of modulating the voice and the Catalan accent* (Spanish, natural stimulus)
#3 *For the microphone pop* (Spanish, natural stimulus)
#4 *Pop sound in the microphone caused by breathing* (Japanese, natural stimulus)

These four comments extracted from the Spanish listeners' perceptual judgments deserve some commentary. Comment #1 and #2 were coded by the authors as "geolect". Geolect, or simply dialect, is the language variety spoken in a particular geographical area. Even though this concept does not rank among the 20 top frequent reasons to classify a stimulus, it is mentioned six times by different Spanish listeners. On all occasions, the classification response is correct. Focusing on the regional variety spoken by the speaker allows listeners to classify correctly at least speakers from Andalusia and Catalonia. This speaks of the lack of dialectal diversity that most cloning algorithms can achieve.

While it is true that there were voices of celebrities, and the listeners reported knowing them beforehand, we should not forget that most times deepfakes are targeted towards these famous people. Given the high hit rate achieved by listeners when paying attention to this high-level linguistic aspect, it seems a good clue for deepfake spotting.

Comment #3 and #4 were coded by the authors as "clipping". Audio clipping is a type of distortion that happens when an audio signal exceeds the maximum level that a recording, amplification, or playback system can handle. This results in harsh, distorted sound, often perceived as buzzing, crackling, or fuzziness. This phenomenon is not uncommon in natural recording settings, even high quality ones, as in interviews humans sometimes talk too loudly or get closer to the microphone for a number of reasons. It seems that this phenomenon is present in, at least, one stimulus in Spanish and one in Japanese. Interestingly, two different listeners paid attention to this aspect and considered it key to classify the stimuli —correctly — as natural. Another similar non-linguistic feature present in the recording that helped listeners correctly identify a stimulus as natural was background noise, which is mentioned five times by listeners. On all occasions, there is a correct audio classification.

# 4. Discussion

## 4.1. Quantitative analysis

Overall participants' performance (59.11%) lies between previously reported results: lower than Müller et al. (2022) and Mai et al. (2023) (approx. 70%), but slightly higher than Cooke et al. (2025) (53.7%), whose experiment was conducted under more realistic, "in the wild", conditions. These findings support the idea that detection accuracy decreases in less controlled environments. Thus, our results align with prior research in suggesting that individuals are highly vulnerable to deceptive synthetic voices and that human perception alone is no longer a reliable safeguard.

This experiment also examined how stimulus characteristics influence detection performance. The effects of language (H1) and speaking style (H2) were confirmed in certain conditions, but the overall pattern was not homogeneous across listener groups, which complicates the formulation of general predictions.

Regarding stimulus language, whenever this variable reached significance, both groups of participants were more accurate when classifying stimuli in their native language. This finding supports the hypothesis that language familiarity enhances deepfake detection (H1) and is consistent with Cooke et al. (2025) and Müller et al. (2022). Furthermore, the effect was stronger among Japanese participants, as it reached significance in more conditions and yielded larger coefficients ($\beta = 1.06$ and $\beta = 1.25$) compared to Spanish participants ($\beta = 0.77$) (Tables 4 and 5).

These findings indicate greater vulnerability to deepfakes presented in a foreign language. Beyond this, the results may also have implications for the selection of evaluators in Text-To-Speech (TTS) assessments. Janska & Clark (2010), for instance, showed that non-native speakers are generally more tolerant of prosodic errors when evaluating a synthetic voice, whereas native speakers prefer flat intonation over defective prosody. Similarly, our results suggest that native and non-native listeners perceive deepfakes differently, which may substantially affect judgments of their quality and naturalness.

With respect to speaking style, significant effects emerged only for stimuli in the native language. Spanish participants classified interviews more accurately than audiobooks, thereby confirming the second hypothesis of the study. This can be explained by the scripted and less spontaneous nature of audiobooks, which contain fewer prosodic irregularities that are difficult to replicate. As a result, TTS algorithms are likely to produce more convincing clones of audiobooks than interviews. This interpretation aligns with research highlighting prosodic modeling as a persistent challenge for TTS systems (Yi et al., 2023). Such limitations make synthetic voices more detectable in spontaneous speech, whereas structured speech styles reduce perceptible mismatches.

In contrast, no consistent pattern was found among Japanese participants. Interviews facilitated performance for natural stimuli, whereas audiobooks facilitated performance for artificial stimuli. As shown in Table 3, mean accuracy for Japanese artificial interviews was notably low (48.85%) compared to artificial audiobooks (65.38%). A plausible explanation for this unexpected result is that the quality of the voice cloning may have been higher than expected, even in spontaneous-speech contexts that are typically considered difficult to replicate. This would suggest that current

TTS systems are advancing rapidly, producing highly realistic voices even under challenging conditions.

These findings also point to an interaction between language and speaking style. This interaction may be explained by the perception of speech disfluencies, which are strongly associated with spontaneous speech. Although disfluencies are a language-universal feature, their realization differs across languages in terms of frequency, distribution, and pragmatic function (Beradze & Meir, 2024). Prior studies have reported difficulties in recognizing and processing disfluencies in a foreign language; for example, non-native listeners sometimes misinterpret disfluencies as parts of words or conversely misinterpret parts of words as disfluencies (Watanabe et al., 2008). Building on this evidence, our results suggest that the facilitative effect of spontaneous style is restricted to the native language, where listeners have the perceptual and pragmatic competence to interpret disfluencies accurately.

These results have broader implications for the development of deepfake detection systems. The performance of such systems depends heavily on the quality and realism of training datasets. Widely used corpora, such as VCTK or LibriSpeech, consist exclusively of audiobook or read-speech recordings. Our results emphasize the importance of speaking style, showing that artificial voices are perceived differently depending on whether the input speech is scripted or spontaneous. Consequently, training datasets should not be limited to reading speech, since systems trained only on scripted materials are likely to yield fragile models that perform poorly in real-world conditions.

Additionally, as in Cooke et al. (2025), both Spanish and Japanese participants in our study were significantly more accurate at identifying authentic than synthetic voices, suggesting a bias toward classifying clips as real. The same finding was observed in Warren et al. (2024), whose study shows that listeners tend to judge audio as human, as reflected in false negative rates being consistently higher than false positive rates in a large-scale study. Given that in most experimental studies participants are explicitly informed about the presence of deepfakes—a condition not typically present outside the lab—, the actual false negative rate in real-world scenarios may be even higher, highlighting the societal risks posed by AI-generated voices. Moreover, as with the variable "language", the variable "authenticity" affected Japanese listeners more strongly than Spanish listeners, since it was significant in more conditions. This explains that Japanese participants exhibited more extreme accuracy rates than Spanish participants (Table 3).

Finally, unlike the findings reported in Wenger et al. (2021), the present study did not find a significant effect of voice familiarity on participants' ability to detect deepfakes. Given the novelty of this research area, further investigation is required. Future studies could explore different degrees of familiarity, including synthetic versions of participants' own voices, as well as voices of acquaintances, friends, and public figures. Studying celebrity deepfakes could provide insights into our vulnerability to fake news, election manipulation, or defamation campaigns targeting public figures. On the other hand, investigating the perception of deepfakes involving acquaintances may be particularly relevant for understanding susceptibility to scams such as vishing.

The quantitative findings discussed so far illustrate the extent to which humans can classify audio in real or artificial, and underscore the need for deeper examination of the factors that shape their judgments. Thus, the second part of this investigation aimed at analyzing the qualitative responses provided by listeners when asked “what information do you pay attention to when deciding whether a voice is natural or artificial?”

## 4.2. Qualitative analysis

Our qualitative analysis offers valuable insights into the strategies humans employ when distinguishing between authentic and synthetic audio. By reducing qualitative responses to their most salient phonetic descriptors, our analysis revealed both common perceptual cues across Japanese and Spanish listeners as well as notable differences in how they conceptualize the “humanness” of speech.

Overall, both Japanese and Spanish listeners relied primarily on suprasegmental and higher-level/non-linguistic features rather than on segmental information to guide their decisions. Concepts such as *intonation, rhythm, fluency, pause, speed, breathing,* and *laughter* were among the most frequently cited in both groups, reinforcing the idea that prosody and disfluency patterns play a central role in human perception of naturalness. These findings are consistent with previous work highlighting the value of prosody and spontaneous speech phenomena in distinguishing human voices from synthetic or cloned speech (Mai et al., 2023; Warren et al., 2024). However, previous studies either provide a list of uncategorized themes (Mai et al. 2023) or propose a thematic classification that lacks the theoretical foundation that in-depth phonetic knowledge provides. In our investigation an effort has been made to classify the most frequent unique words provided by our listeners in the three most common groups typically used in phonetic studies. For instance, breathing, mouth noises and nasal are keywords classified with the code ‘liveliness’ and within the theme “external” in Warren et al. (2024). That classification seems somewhat unconventional and may benefit from stronger theoretical grounding in phonetics. However, we can still find that some of the repeated topics in previous studies are the same as those found in our study; for instance, the topic “prosody”, which in Warren et al. (2024) is composed of keywords “tone”, “inflections”, “cadence”, “pitch”, among others.

Importantly, in our study non-linguistic features such as breathing were repeatedly mentioned as strong indicators of naturalness both by Spanish and by Japanese listeners, suggesting that listeners treat them as hallmarks of genuine human speech production. This finding agrees with many previous studies (Mai et al., 2023, Warren et al., 2024; Layton et al., 2024). This strong focus on respiratory cues that can be observed across studies may reflect a heightened perceptual sensitivity to bodily traces of speech, aligning with the broader observation that listeners often interpret breath as an unmistakably human marker.

Another cross-group convergence concerns emotionality. Both Japanese and Spanish listeners associated *emotion* with the perceived naturalness of the stimuli. Expressions such as “lack of emotion” or “emotionless” were used to characterize artificial voices, aligning with the view that emotional expressivity is one of the most challenging aspects to replicate in synthetic speech.

Despite these shared tendencies, interesting differences emerged between the two listener groups. For Japanese listeners, the most frequent terms were *inflection, intonation,* and *fluency.* Their responses emphasized the dynamic aspects of pitch and amplitude variation as central to judgments of naturalness. While Japanese is not a fully tonal language, it is a pitch-accent language, in which specific syllables are marked by high (H) or low (L) pitch patterns that serve to distinguish word meanings. The predominance of the term *inflection* in the responses likely reflects sensitivity to unnatural pitch-accent stress, where inaccuracies in pitch-accent placement may create perceptual anomalies.

The second most frequent term among Japanese, *intonation,* can be reasonably interpreted as referring to broader prosodic features at the sentence level. However, the distinction between pitch accent and intonation is not always clearly made in lay usage. Indeed, individuals without formal training in Japanese phonology often conflate pitch accent (which differentiates word meanings) with intonation (which conveys sentence-level prosody or emotion). As a result, listeners may describe deepfake speech as having "intonational errors," when what they are in fact perceiving are pitch-accent anomalies. Another interesting aspect worth mentioning is that the technical term *prosody* is not common in everyday Japanese, which explains its absence from participant feedback, in contrast to Spanish listeners who used the word *prosody* explicitly.

In contrast, Spanish listeners displayed a broader range of phonetic references, notably including segmental detail. Thirteen responses explicitly referred to consonants, and in most cases these were linked to correct classifications of stimuli. This reliance on fine-grained phonetic detail is likely related to the relatively high proportion of participants with advanced linguistic training (39 %), a factor absent from the Japanese sample. Furthermore, Spanish listeners uniquely employed more technical terminology such as *prosody* and *timbre,* terms unlikely to appear spontaneously in the responses of non-specialists. The presence of dialectal cues (*geolect*) was another distinctive feature: some participants explicitly referred to Andalusian or Catalan speech, exploiting regional variation as a diagnostic marker. This type of meta-linguistic awareness underscores the role of linguistic expertise and sociophonetic knowledge in shaping perceptual strategies.

The results suggest that both universal perceptual cues (e.g., disfluencies, suprasegmentals, non-linguistic sounds) and culture- or expertise-dependent cues (e.g., segmental detail, dialectal knowledge, pitch-accent anomalies) contribute to the identification of synthetic speech. From an applied perspective, this highlights two important directions for the improvement of deepfake detection algorithms. First, computational models could benefit from incorporating markers of disfluency and non-linguistic vocal events, as these appear to be robust cross-linguistic indicators of naturalness. Second, language-specific phonological systems such as Japanese pitch accent provide another valuable avenue: detection systems could be trained to identify subtle pitch-accent mismatches, which naïve listeners already report as strong indicators of artificiality.

# 5. Limitations and future work

The study is not without limitations. First, the results do not fully reflect human detection performance as it would be in the wild. Our experiment did not fully replicate real-world conditions: listeners were explicitly informed that some of the voices were artificial, and the sentences they listened to were presented without contextual cues that, in natural settings, might aid discrimination.

A further limitation concerns uncontrolled variables. In addition to the factors we examined, other factors may have influenced performance. These include distractions, participant fatigue, ambient noise, and difficulties using the online platform. Because the study was conducted remotely and without direct supervision, the influence of such uncontrolled factors may have been magnified.

Another issue relates to the variable voice familiarity. Participants self-reported whether they recognized the celebrities' voices, but no objective test was used to confirm their familiarity. A more reliable design would incorporate a brief voice-recognition task to verify the accuracy of these self-reports.

Furthermore, a key limitation of our study concerns the reduction of qualitative responses into single words, which, while necessary for systematic analysis, inevitably abstracts away from the richness of listeners' judgments. As shown in the quoted sentences, full responses often contain nuanced observations (e.g., about microphone clipping or discourse markers) that resist one-word categorization but may be crucial for understanding perceptual decision-making. Future research should therefore combine quantitative word-based analyses with more fine-grained qualitative coding.

Finally, the asymmetry in participant profiles—particularly the higher proportion of linguistically trained listeners among the Spanish participants—suggests caution when comparing groups. While the findings reveal genuine differences in perceptual strategies, some of these may be attributable to educational background rather than to cross-cultural variation per se. Further studies with more balanced samples would be required to disentangle cultural from expertise-driven effects.

Other avenues for future research emerge from the present study. (a) It would be valuable to evaluate the in-domain and out-of-domain performance of an algorithm trained on the voices used in this experiment, and to compare its detection accuracy with that of human participants, following a similar approach to that of Mai et al. (2023). (b) Further exploration of reaction times in deepfake discrimination, as proposed by San Segundo & Gibson (2024), could clarify whether speed correlates with accuracy. (c) Beyond perceptual cues, examining the acoustic features of natural and synthetic stimuli—as in San Segundo & Delgado (2024)—could reveal measurable indicators linked to classification accuracy.

# 6. Conclusions

Audio deepfakes have emerged as a growing global concern, extending beyond the security domain to broader society. This study investigates human performance in classifying deepfake audio and the factors shaping their decisions. To this end, we conducted an online user study in which Spanish and Japanese listeners classified samples of different speaking styles (interviews vs. text reading). Results suggest an average accuracy of 59.11%, with higher performance on authentic samples. The study also demonstrates that judgments of vocal naturalness are grounded in a combination of linguistic and non-linguistic cues. The latter appear to function as universal perceptual cues, while linguistic expertise and language-specific phonological information enable additional reliance on segmental or structural information. These results underscore the complexity of human perceptual strategies in distinguishing natural from artificial speech and provide insight into human perceptual strategies.